\documentclass[a4paper,twoside]{article}

\usepackage{epsfig}
\usepackage{subcaption}
\usepackage{calc}
\usepackage{amssymb}
\usepackage{amstext}
\usepackage{amsmath}
\usepackage{amsthm}
\usepackage{multicol}
\usepackage{pslatex}

\usepackage{latexsym}

\usepackage{algorithm2e}
\SetKwFunction{cylinder}{cylinder}
\SetKwFunction{smoothstep}{smoothstep}
\SetKwFunction{saturate}{saturate}

\usepackage{color,soul}

\usepackage{multirow}

\usepackage{array}

\usepackage{float}

\usepackage{hyperref}

\usepackage{natbib}
\let\bibhang\relax
\usepackage{apalike}

\usepackage{SCITEPRESS}     

\begin{document}

\title{Hybrid MBlur: A Systematic Approach to Augment Rasterization with Ray Tracing for Rendering Motion Blur in Games}
\author{\authorname{Anonymous}}

\author{\authorname{Yu Wei Tan\orcidAuthor{0000-0002-7972-2828}, Xiaohan Cui\orcidAuthor{0000-0002-6152-1665} and Anand Bhojan\orcidAuthor{0000-0001-8105-1739}}
\affiliation{School of Computing, National University of Singapore, Singapore}
\email{\{yuwei, cuixiaohan\}@u.nus.edu, banand@comp.nus.edu.sg}
}

\keywords{Real-time, Motion Blur, Ray Tracing, Post-processing, Hybrid Rendering, Games.}

\abstract{Motion blur is commonly used in game cinematics to achieve photorealism by modelling the behaviour of the camera shutter and simulating its effect associated with the relative motion of scene objects. A common real-time post-process approach is spatial sampling, where the directional blur of a moving object is rendered by integrating its colour based on velocity information within a single frame. However, such screen space approaches typically cannot produce accurate partial occlusion semi-transparencies. Our real-time hybrid rendering technique leverages hardware-accelerated ray tracing to correct post-process partial occlusion artifacts by advancing rays recursively into the scene to retrieve background information for motion-blurred regions, with reasonable additional performance cost for rendering game contents. We extend our previous work with details on the design, implementation, and future work of the technique as well as performance comparisons with post-processing.}

\onecolumn \maketitle \normalsize \setcounter{footnote}{0} \vfill

\section{\uppercase{Introduction}}

We showcase a novel real-time hybrid rendering technique for the Motion Blur (MBlur) effect that combines ray tracing and post-processing, exploiting ray-traced information within a selection mask to reduce partial occlusion artifacts in post-processed MBlur. We provide a thorough illustration and evaluation of hybrid MBlur, extending our published work \citep{Tan:2020:HMB} with the following additions:
\begin{itemize}
	\item Extensive discussion of the design and implementation details of hybrid MBlur.
	\item Visual quality and performance evaluations of our technique in comparison to post-processing.
	\item Explanation of presented extensions and future work to the approach.
\end{itemize}

\subsection{Background Information}

\subsubsection{Hybrid Rendering}

Ray tracing is a common approach to produce realistic effects like glossy reflections, depth of field, and motion blur. However, its high computation time has limited its use mostly to offline rendering. On the other hand, many post-process techniques have been devised to approximate these effects on rasterized images to meet the constraint of interactive frame rates for games, albeit with limited visual quality. 

Given recent developments in hardware acceleration, it is now the time to marry ray tracing with rasterization for games. Ray tracing can be employed to correct certain visual flaws originating from pure post-process approaches, while still adhering to the performance budgets of real-time rendering. This results in more realistic and convincing graphics while keeping the gaming experience interactive, enhancing the overall immersion of the player in the game. 

\subsubsection{MBlur}

\begin{figure}[!h]
	\centering
	\includegraphics[width=\linewidth]{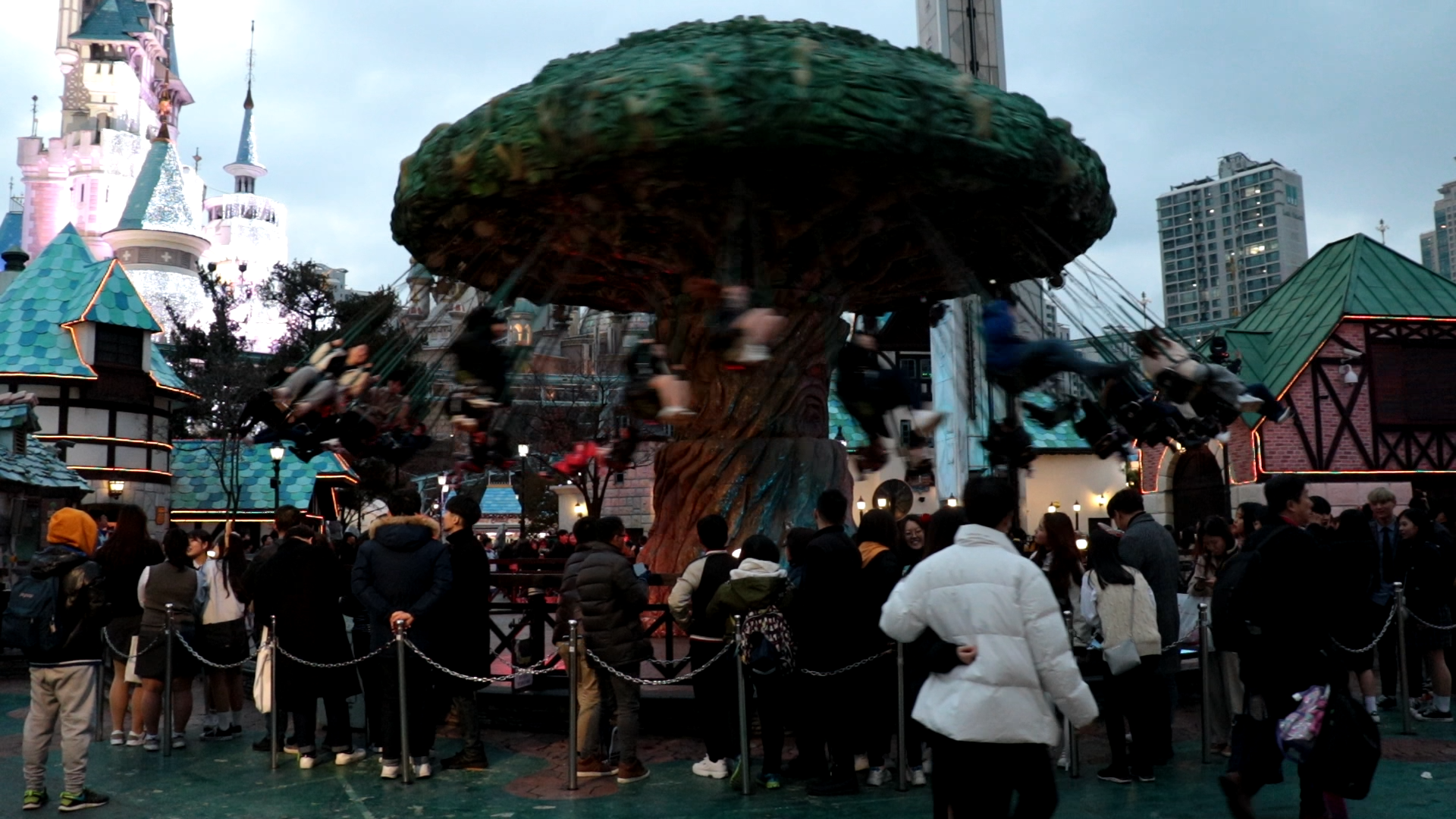}
	\caption{MBlur effect. Image from authors.}
	\label{fig:mb-example} 
\end{figure}

MBlur is the streaking or smearing effect of objects in the direction of relative motion with respect to the camera. It is employed in cinematics to emphasize the relative speed of moving objects, producing a blurring effect as shown in \autoref{fig:mb-example}. The MBlur effect has been used to express the speed of rapid movement, the disoriented state of mind of a character, or the dreamlike quality of a scene. 

In cameras, MBlur is produced with the choice of exposure time or shutter speed, which is the duration for which the camera shutter is open per second. This duration directly correlates to the amount of time the camera sensor is exposed to light. Hence, when the exposure time is long, every resulting image does not represent an instantaneous moment but rather, an interval consisting of successive moments in time. An object moving in relation to the camera will therefore appear on multiple areas of the sensor over time, resulting in a directional blur effect of the object corresponding to its motion. 

As described in \citet{Jimenez:2014:ARR}, moving objects blur both outwards and inwards at their silhouettes, making the region around their silhouettes appear semi-transparent. Outer blur represents an object's blur into its neighbouring background while inner blur applies to the blur produced within the silhouette of the object itself. According to \citet{Cook:1984:DRT}, accurate motion blurring takes into account areas of background geometry occluded by any blurred foreground. Hence, realistic MBlur should have accurate partial occlusion by recovering true background colour in inner blur regions instead of approximating this colour like many post-process approaches. Background colour recovery in MBlur also helps to prevent inaccuracies between real and approximated backgrounds for sharp and blurred regions respectively. Hence, we employ ray tracing in our approach to retrieve the exact colour of occluded background.
\section{\uppercase{Related Work}}

\subsection{Hybrid Rendering}
 
The use of ray tracing to complement rasterization techniques to achieve high graphics quality while maintaining interactive frame rates has been attempted for several visual effects over the years. In \citet{Beck:1981:HGR} and \citet{Hertel:2009:HGR}, shadow mapping is employed to mark shadow boundaries and reduce the actual number of shadow rays to be traced, of which only rays from regions deemed potentially inaccurate by shadow mapping are traced in \citet{Lau:2009:FHS} to generate precise and alias-free shadows. For realistic reflections, \citet{Macedo:2018:FRR} invokes ray tracing only on pixels with reflections that cannot be solved with just screen space information. In \citet{Marrs:2018:ATA}, traditional temporal anti-aliasing (TAA) is also extended with ray-traced supersampling for regions with a high chance of TAA failure. 

In \citet{Cabeleira:2010:CRR}, diffuse illumination is computed via rasterization while reflections and refractions are handled by ray tracing. In more general pipeline approaches, \citet{Chen:2007:UHZ} substitutes the generation of primary rays in recursive ray tracing \citep{Whitted:1979:IIM} with rasterization for improved performance. This approach is extended in \citet{Andrade:2014:THB} which adheres to a render time limit by tracing rays only for objects of the highest importance.

\subsection{MBlur}

Current implementations of MBlur include the use of an accumulation buffer. Within the exposure time when the camera shutter is set to be open, a series of discrete snapshots rendered at multiple time offsets from the actual time value of the image is integrated to produce MBlur such as in \citet{Korein:1983:TAC} and \citet{Haeberli:1990:ABH}. However, as discussed in \citet{McGuire:2012:RFP}, this approach is computationally expensive in terms of shading time. Furthermore, in the case of undersampling, this method can lead to a series of disjoint ghosting artifacts in the direction of motion instead of a continuous blur.


Stochastic sampling \citep{Cook:1986:SSC} for MBlur is a Monte Carlo technique based on \citet{Halton:1970:RPS} in which sampling is performed in a randomized and nonuniform manner along the time dimension. \citet{McGuire:2010:RSR} incorporates stochastic sampling with ray tracing by conservatively estimating the total convex hull bound for the area of each moving triangle during the exposure time, leveraging hardware rasterization. Stochastically distributed rays are then shot into the scene within this bound for shading. This approach makes use of distributed ray tracing \citep{Cook:1984:DRT}, where rays are distributed in time and the colour at each hit point is averaged to produce the final image. 

\citet{Cook:1987:RIR} translates micropolygons for each sample according to a jittered time. On the other hand, \citet{Fatahalian:2009:DRM} and \citet{Sattlecker:2015:RRG} achieve efficiency in performance by extending the bounding box of each micropolygon to consider its pixels and their associated velocities instead. \citet{Hou:2010:MRT} is an approach that makes use of 4D hyper-trapezoids to perform micropolygon ray tracing. Methods of approximating the visibility function to sample have also been devised such as in \citet{Sung:2002:STA}. However, as explained in \citet{Sattlecker:2015:RRG}, at low sample rates, these methods will also exhibit ghosting artifacts whereas increasing the number of samples would lead to noise.

To produce the right amount of blur, some real-time approaches like \citet{Rosado:2008:GGP}, \citet{Ritchie:2010:SSM} and \citet{Sousa:2013:GGF} make use of per-pixel velocity information by accumulating samples along the magnitude and direction of velocities in the colour buffer. Other techniques in \citet{Korein:1983:TAC}, \citet{Catmull:1984:AVS} and \citet{Choi:2017:RMB} accumulate the colours of visible passing geometry or pixels with respect to a particular screen space position while \citet{Gribel:2011:HSR} makes use of screen space line samples instead. \citet{Potmesil:1983:MMB} represents the relationship between objects and their corresponding image points as point-spread functions (PSFs), which are then used to convolve points in motion. \citet{Leimkuhler:2018:LKS} splats the PSF of every pixel in an accelerated fashion using sparse representations of their Laplacians instead. Time-dependent edge equations, as explained in \citet{Akenine:2007:SRT} and \citet{Gribel:2010:AMB}, and 4D polyhedra primitives \citep{Grant:1985:IAS} have also been used for MBlur geometry processing. Recently, a shading rate-based approach involving content and motion-adaptive shading in \citet{Yang:2019:VLC} has also been developed for the generation of MBlur.

In particular, attempts to simulate nonlinear MBlur include \citet{Gribel:2013:TAH} and \citet{Woop:2017:STB}. Our hybrid technique only considers linear inter-frame image space motion for now, but we intend to provide support for higher-order geometry motion in the future. We also assume mainstream ray tracing acceleration architecture widely available in modern gaming workstations. The GA10x RT Core of the newest NVIDIA Ampere architecture provides hardware acceleration for ray-traced motion blur \citep{NVIDIA:2021:NAG} but is only found in the premium GeForce RTX 30 Series graphics cards.

\section{\uppercase{Design}}

\begin{figure*}[!h]
	\centering
	\includegraphics[height=0.85\textheight,keepaspectratio]{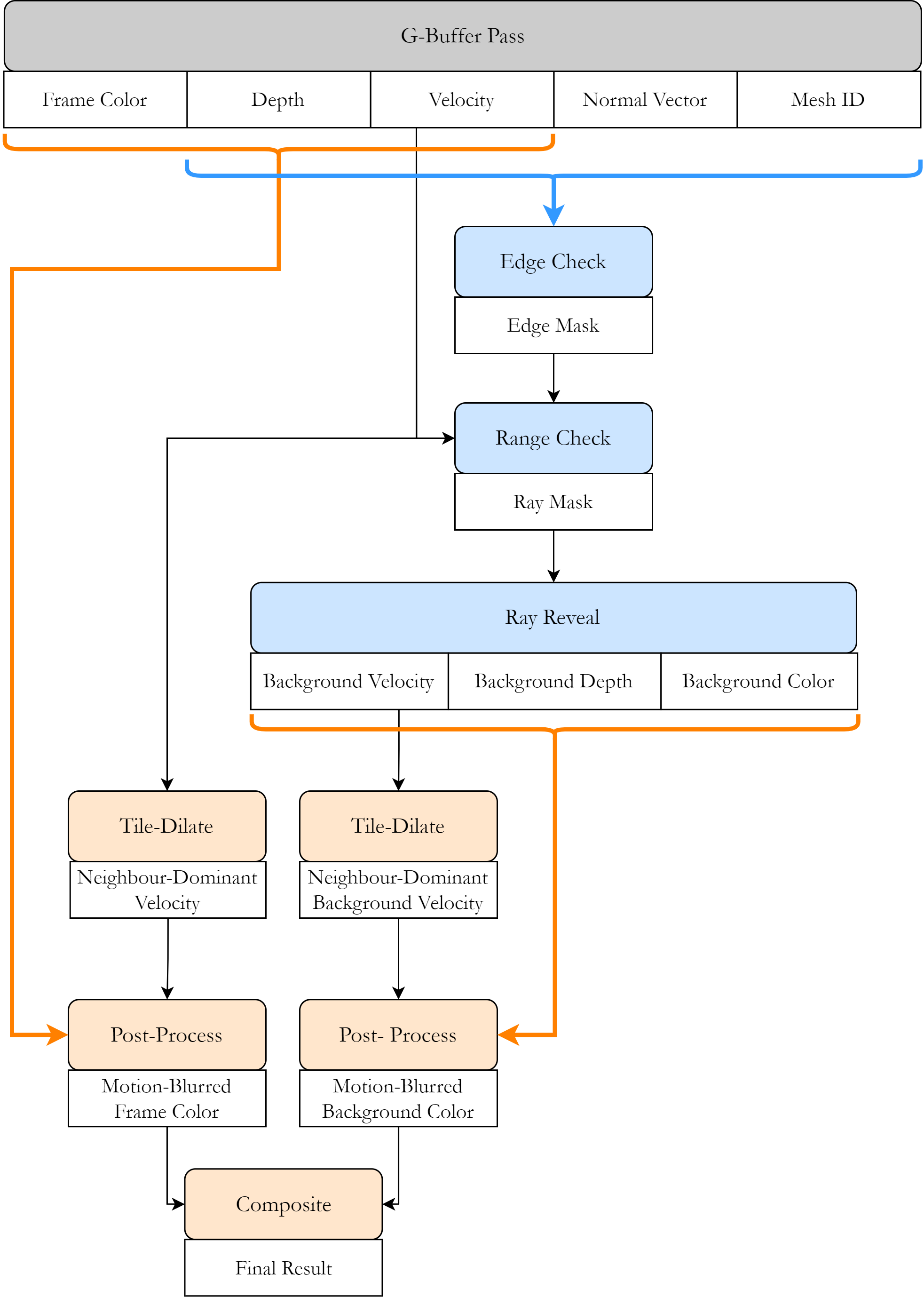}
	\caption{Hybrid MBlur rendering pipeline.}
	\label{fig:mb-pipeline}
\end{figure*}

Our hybrid MBlur approach, as illustrated in \autoref{fig:mb-pipeline}, compensates for missing information in post-processed MBlur with the revealed background produced by a ray trace-based technique. 

A Geometry Buffer (G-Buffer) is first generated under a deferred shading set-up, rendering textures containing per-pixel information such as camera space depth, screen space velocity and rasterized colour. The same depth, velocity and colour information for background geometry is produced by our novel ray reveal pass within a ray mask for pixels in the inner blur of moving foreground objects. A tile-dilate pass is then applied to these 2 sets of buffers to determine the sampling range of our gathering filter in the subsequent post-process pass. Both the ray-revealed and rasterized output are blurred by this post-process pass and lastly composited together.

\subsection{Post-process}

The \citet{McGuire:2012:RFP} post-process MBlur renders each pixel by gathering sample contributions from a heuristic range of nearby pixels. We adapt this approach to produce a motion-blurred effect separately for rasterized and ray-revealed information.

As suggested in \citet{Rosado:2008:GGP}, motion vectors are given by first calculating per-pixel world space positional differences between every frame and its previous frame, followed by a translation to screen space. By using the motion vector between the last frame and the current frame, we simulate the exposure time of one frame. Then, we follow the approach of \citet{McGuire:2012:RFP} in calculating the displacement of the pixel within the exposure time by scaling the inter-frame motion vector with the frame rate of the previous frame as well as the exposure time. Considering the full exposure as one unit of time, this displacement can be interpreted as a per-exposure velocity vector. This approach uses the assumption that the motion vector of each pixel between the previous frame and the current frame remains constant throughout the exposure time. 

\citet{Jimenez:2014:ARR} is a technique that is based on \citet{McGuire:2012:RFP}. As described in \citet{Jimenez:2014:ARR}, the major problems to be addressed when producing a post-processed MBlur are the range of sampling, the amount of contribution of each sample as well as the recovery of background geometry information for inner blur. With our method for calculating per-exposure velocities, we adopt \citet{McGuire:2012:RFP}'s approach in determining the magnitude of the sampling range and representing different amounts of sample contribution, as illustrated in the Appendix for completeness. As shown in \autoref{fig:mb-samplerange}, \citet{McGuire:2012:RFP} centers the sampling area at the target pixel, creating a blur effect both outwards and inwards from the edge of the object. Although this produces a more uniform blur for thin objects and the specular highlights of curved surfaces, it poses the problem of having to smoothen the transition between the inner and outer blur. Hence, we let the pixel be at the end of the sampling area instead, so inner and outer blur can be considered separately without much change to the pipeline. We also enhance the inner blur with ray-revealed background information.

\begin{figure}[!h]
	\centering
	\includegraphics[width=\linewidth]{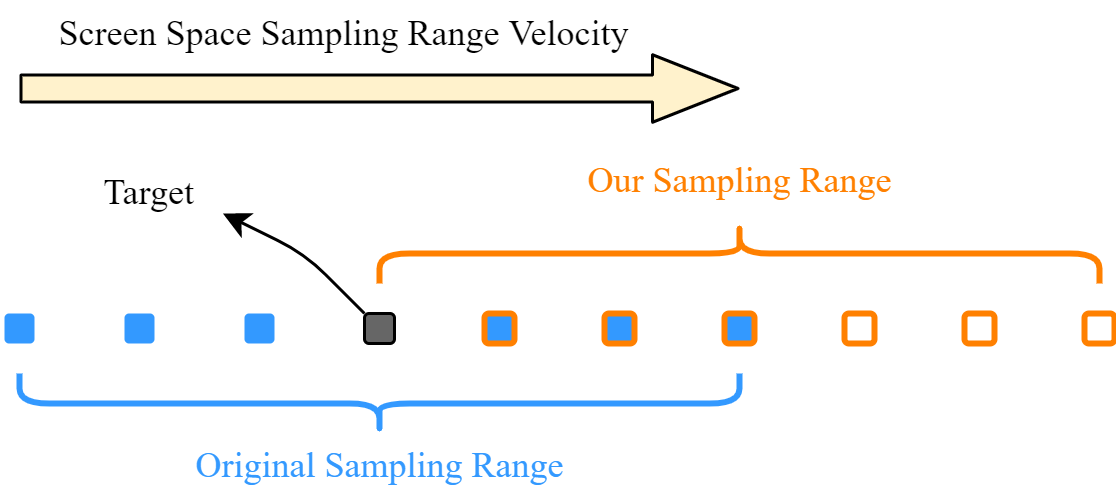}
	\caption{Range of samples for each pixel.}
	\label{fig:mb-samplerange}
\end{figure}

\subsection{Ray Reveal}

\citet{McGuire:2012:RFP} approximates the inner blur region of objects with colour information of nearby samples. This is based on the reasoning that approximate background information is still better than the absence of any background. With our ray reveal algorithm, we ray trace to obtain true background information for the inner blur.

\subsubsection{Ray Mask}

We adopt selective rendering \citep{Chalmers:2006:SRC} in generating a ray mask that determines regions of inner blur corresponding to geometry edges and only shoot rays within the mask for better performance like in adaptive frameless rendering \citep{Dayal:2005:AFR}. 

For every pixel with a nonzero speed, we first translate it by its velocity for one magnitude of its estimated displacement within the exposure time to predict its next screen space position at the end of the exposure duration and compare their mesh ID and depth. A target pixel is filtered out of the mask if the following conditions are not met:
\begin{equation}
m_{n} \neq m_{t}
\end{equation}
\begin{equation}
z_{n} - z_{t} > {\rm SOFT\_Z\_EXTENT}
\end{equation}

$m_{n}$ and $m_{t}$ refer to the corresponding mesh ID of the pixel's next and current positions respectively. We assume that every object has uniform speed throughout its geometry, so we reject the case when the $m_{n}$ and $m_{t}$ are equal. $z_{n}$ and $z_{t}$ refer to the camera space depth of the pixel's next and current position respectively. For a target pixel to be in the inner blur region of a foreground object, it would have to be shallower than its next position. Hence, $z_{n}$ has to be deeper than $z_{t}$ by a value greater than SOFT\_Z\_EXTENT, a scene-dependent variable with a positive value as introduced in the \citet{McGuire:2012:RFP} paper for calculating sample contribution.

For pixels that are not filtered out, they are passed through a $5 \times 5$ Sobel convolution kernel which estimates how extreme an edge is. The kernel is applied to the G-Buffer to obtain an approximate derivative of the gradient associated with each pixel, based on the depth and surface normal information of its vicinity readily accessible from the deferred shading stage. The depth derivatives identify the divide between overlapping objects where the colour of one object might be revealed through the other, while the normal derivatives can locate pronounced differences in the orientation of primitive faces within objects themselves near their silhouettes. The output of each pixel from this filter is hence computed as follows:
\begin{equation}
    x = \delta_{d}+ \delta_{n}
\end{equation}
\begin{equation}
    x_{n} = \saturate{$1 - \frac{1}{x + 1}$}
\end{equation}

Here, $\delta_{d}$ and $\delta_{n}$ refer to the depth derivative and length of normal derivatives respectively from the Sobel operator. The normalized $x$, $x_{n}$, is subsequently evaluated against an edge threshold $e$, which is set to be high in order to effectively eliminate non-edges from the mask. Only pixels that pass the threshold test will be marked in the edge mask. 

Lastly, a range check pass is applied. Each pixel with a nonzero speed then samples along the direction of its velocity. If any of its samples is marked in the edge mask, the pixel passes into the final ray mask. 

\subsubsection{Recursive Ray Tracing}

\begin{figure}[!h]
	\centering
	\includegraphics[width=\linewidth]{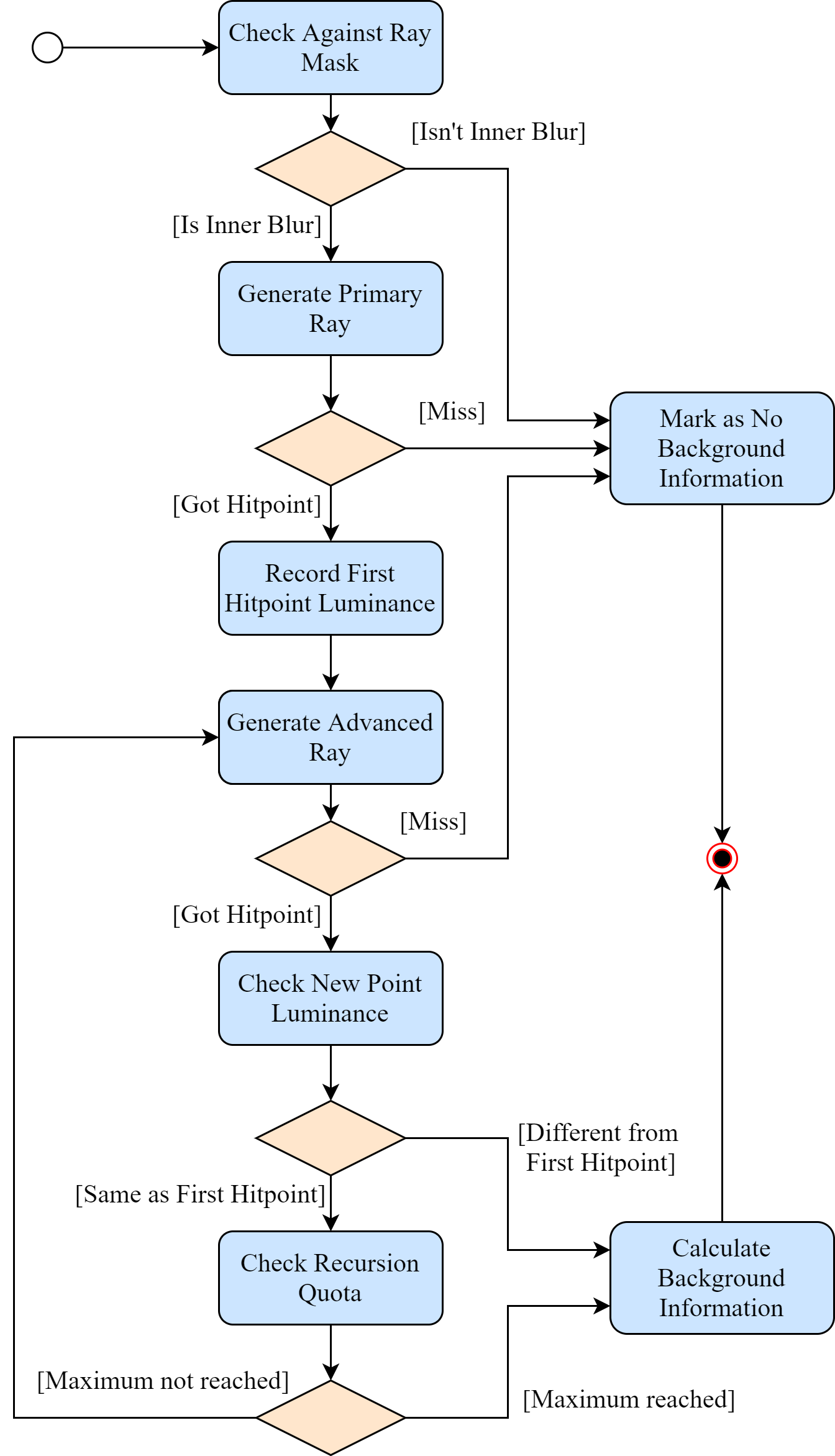}
	\caption{Ray tracing approach.}
	\label{fig:mb-rayreveal}
\end{figure}

To obtain more accurate information behind foreground objects for compositing inner blur, our ray tracing approach, as illustrated in \autoref{fig:mb-rayreveal}, adapts the idea of recursive ray tracing \citep{Whitted:1979:IIM} where we shoot rays and iteratively advance them deeper into the scene until a different object is found. At the end of the recursion, the occluded background is then revealed with the final hit point. Revealing more layers deeper into the scene with this method increases the accuracy of our motion-blurred background but at diminishing returns. Hence, we limit the ray reveal process to only one background layer and post-process (i.e. approximate using neighbour information) it instead.

For every pixel marked in the ray mask, we shoot a ray from the camera into its world space position and immediately spawn a second ray along the same direction after the first hit. Currently, we use a simple indicator, the difference in luminance, to identify different objects. Hence, we terminate the recursion when the latest hit point reads a luminance different from the first hit or the maximum number of recursions is reached.
\section{\uppercase{Implementation}}

To implement our hybrid MBlur approach, we made use of the NVIDIA Falcor real-time rendering framework \citep{NVIDIA:2017:FRF} on DirectX 12 for ray tracing acceleration. The scenes used for testing our approach are \textsc{The Modern Living Room} (commonly referred to as \textsc{Pink Room}) \citep{Wig42:2014:MLR}, \textsc{UE4 Sun Temple} \citep{EpicGames:2017:UES} and the interior scene of \textsc{Amazon Lumberyard Bistro} (\textsc{Bistro Interior}) \citep{AmazonLumberyard:2017:ALB}.

For our post-process tile-dilate pass, we followed \citet{McGuire:2012:RFP}'s approach of taking a neighbourhood size of $n = 3$ but found that a tile size of $m = 40$ generally worked better than $m = 20$ for us in our test scenes with fast-moving foreground objects. This is because the tile length scales the amount of blur we can observe in the final image, and we require a substantial amount of blur to expose and inspect the quality of semi-transparencies of foreground silhouettes through which background geometry is revealed. Hence, we also tested our scenes at long exposure times or low shutter speeds of $1 / 60$ s, against high frame rates of 100 to 300 fps so as to scale our per-exposure velocities effectively. As for our sample count, we took 15 samples along the magnitude of the dominant neighbourhood velocity as recommended in \citet{McGuire:2012:RFP}. Our SOFT\_Z\_EXTENT was 3 cm (\textsc{Pink Room} and \textsc{Sun Temple}) and 5 mm (\textsc{Bistro Interior}), which are also within the suggested range of values from \citet{McGuire:2012:RFP}.

As for our ray mask, we employed a strict edge threshold $e$ of 0.9 (\textsc{Pink Room}), 0.98 (\textsc{Sun Temple}) and 0.96 (\textsc{Bistro Interior}) which were effective in detecting the intricate edges of scene objects while rejecting non-edges well. For the ray tracing pass, we limited the recursion level to 5. However, for complex scenes with concave objects that lead to a significant amount of self-occlusion, this level can be increased.

When the target pixel is deeper than its sample by at least SOFT\_Z\_EXTENT in the post-process pass, we artificially magnify $w_{f}$ by 30 for a smooth colour transition from the object's edge to its outer blur. On the other hand, we deem the pixel to be in the inner blur region based on whether it is marked in the ray mask. If the pixel is in the mask, we also optionally magnify its overall background colour weight ($w$ in background colour) and diminish its foreground colour weight ($w$ in foreground colour) accordingly by the same magnitude of 3, so background geometry can be observed more clearly through the silhouette of foreground objects with partial occlusion.

\section{\uppercase{Results}}

Our hybrid MBlur approach (d) aims to improve the visual quality of the post-process technique in \citet{McGuire:2012:RFP} with reasonable additional performance cost. Besides the original rasterized colour (a) and an adapted version of the post-process technique at the new sampling range as shown in \autoref{fig:mb-samplerange} (b), we also evaluate our approach against state-of-the-art post-process MBlur from Unreal Engine 4 (UE4) (c), which adopts a similar technique to \citet{McGuire:2012:RFP}. UE4 also produces MBlur by comparing the velocity of each pixel to that of its neighbours \citep{EpicGames:2020:UED}. The ground truth is given by the distributed ray tracing approach from \cite{Cook:1984:DRT} (e). Our demo video (\href{https://youtu.be/k3mgDjDEe_A}{link}) provides more detailed comparisons of the results.

\begin{figure*}[!h]
	\centering
	\subcaptionbox{}{
		\includegraphics[width=0.18\linewidth]{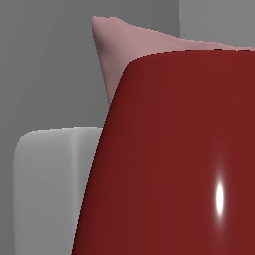}
	}
	\subcaptionbox{}{
		\includegraphics[width=0.18\linewidth]{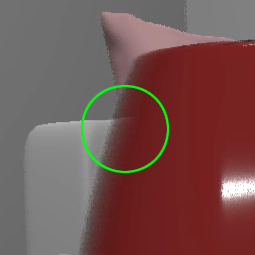}
	}
	\subcaptionbox{}{
		\includegraphics[width=0.18\linewidth]{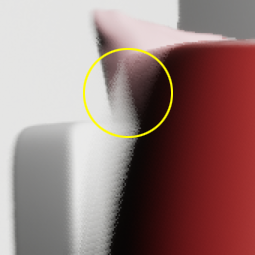}
	}
	\subcaptionbox{}{
		\includegraphics[width=0.18\linewidth]{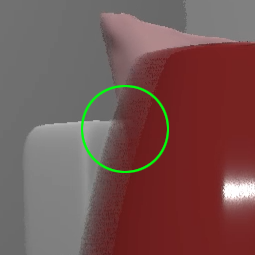}
	}
	\subcaptionbox{}{
		\includegraphics[width=0.18\linewidth]{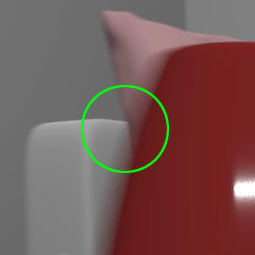}
	}
	\par\smallskip
	\caption{Comparison of partial occlusion quality.}
	\label{fig:mb-partialocclusion}
\end{figure*}

\subsection{Graphics Quality Comparison}

Post-process MBlur inaccurately reconstructs the background by reusing neighbouring information available in the initial sharp rasterized image, which creates artifacts such as mismatched patterns when the background contributing to the inner blur is not of a uniform colour. In \autoref{fig:mb-partialocclusion}, we choose a shot from \textsc{Pink Room} with a moving red vase in front of a white sofa with a pink cushion on it. For the adapted post-process MBlur from \citet{McGuire:2012:RFP}, the silhouette of the cushion incorrectly terminates at the edge of the vase. While the pink colour from the cushion should extend towards the bottom right direction into the geometry of the vase, it is blocked by a horizontal division created by motion vector-based sampling. This artifact is more noticeable for UE4, where the background colour within the inner blur is practically a linear translation from the left of the edge, as shown in the yellow circle. However, our hybrid method samples what is behind the inner blur of moving objects, hence we can obtain the desired background, which in this case is the extension of the cushion silhouette in the green circle corresponding to the ground truth.

However, it can be seen that hybrid MBlur does not achieve as much semi-transparency as compared to the ground truth, where the background colour is more visible within the silhouette of the vase. Nonetheless, the rendering time of the ground truth is 13 fps at 200 rays per pixel (rpp), which is a lot slower than 205 fps for hybrid MBlur. In practice, it is not common to use distributed ray tracing to render MBlur in real-time as it is difficult to meet interactive frame rates, while reducing the rpp to achieve ideal performance will introduce a significant amount of sampling noise. 

Against the ground truth, our hybrid MBlur method performs well in objective visual metrics PSNR and SSIM when compared to other state-of-the-art post-process methods, as presented in \autoref{tab:metrics}. It is also interesting to note that the improved MBlur method implemented in the latest version of UE4 does not fare as well as the basic adapted post-process method from \citet{McGuire:2012:RFP}.

\begin{table}[!h]
\begin{center}
	\caption{Comparison of similarity to ground truth.}
	\label{tab:metrics}
\begin{tabular}{|l|l|l|}
	\hline
	\textbf{}            & \textbf{PSNR} & \textbf{SSIM} \\ \hline
	Adapted Post-Process & 24.20         & 0.91          \\ \hline
	UE4 Post-Process     & 12.32         & 0.68          \\ \hline
	Hybrid MBlur              & \textbf{25.69}         & \textbf{0.92}          \\ \hline
\end{tabular}
\end{center}
\end{table}
  
\subsection{Performance}
	
\begin{figure*}[!h]
	\centering
	\subcaptionbox{\textsc{Pink Room (PR)}}{
		\includegraphics[width=0.31\linewidth]{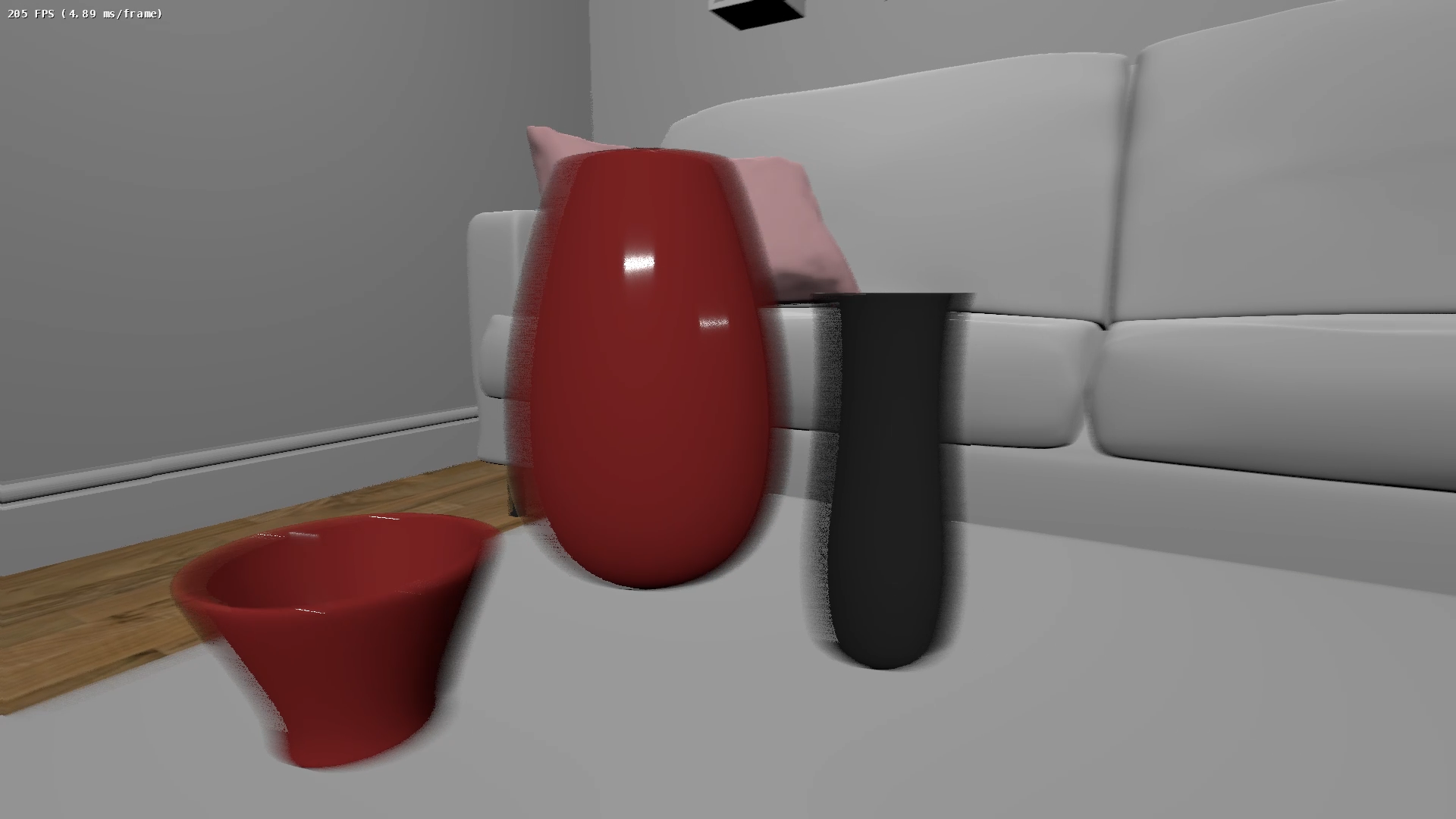}
	}
	\subcaptionbox{\textsc{Sun Temple (ST)}}{
		\includegraphics[width=0.31\linewidth]{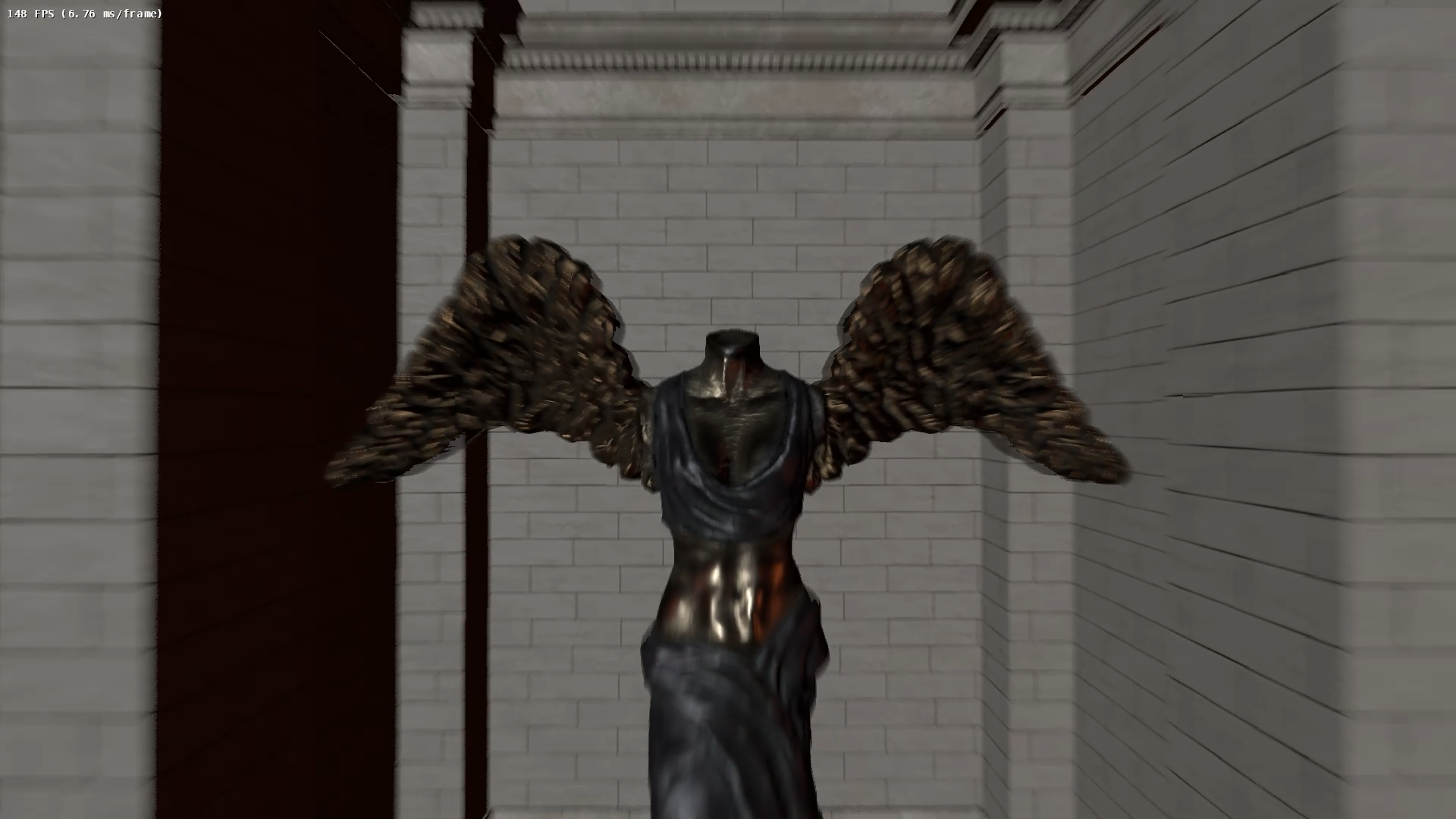}
	}
	\subcaptionbox{\textsc{Bistro Interior (BI)}}{
		\includegraphics[width=0.31\linewidth]{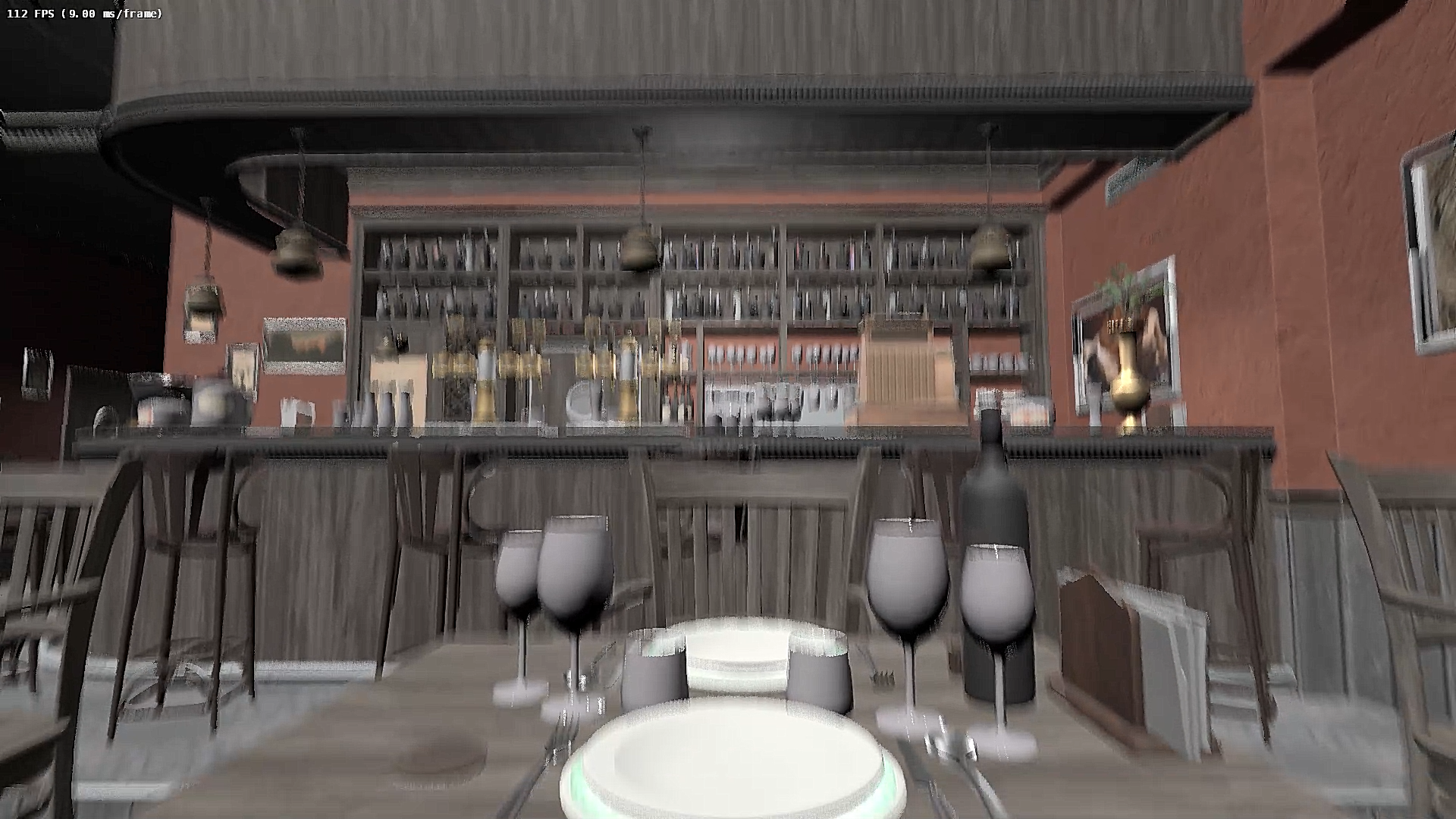}
	}
	\par\smallskip
	\caption{Shots used for profiling.}
	\label{fig:mb-profiling}
\end{figure*}

Although the Falcor framework helped facilitate the implementation of our technique by taking care of scene loading, pipeline set-up and the creation of ray tracing acceleration structures, the abstraction of many low-level details from Direct3D makes it difficult to optimize rendering. However, even without substantial optimization, we have managed to achieve relatively interactive frame rates and pass durations, as shown in \autoref{tab:mb-profiling} based on shots of varying geometric complexity in \autoref{fig:mb-profiling}. Our measurements are taken with the Falcor profiling tool on an Intel Core i7-8700K CPU at 16GB RAM with an NVIDIA GeForce RTX 2080 Ti GPU.

\begin{table}[!h]
\begin{center}
	\caption{Pass durations (in ms) and frame rates.}
	\label{tab:mb-profiling}
\begin{tabular}{|l|c|c|c|c|}
	\hline
	Shot                                 & Processor    & \textsc{PR} & \textsc{ST} & \textsc{BI} \\ \hline
	                                     & CPU          & 0.26               & 0.54                & 1.74                      \\ \cline{2-5}
	\multirow{-2}{*}{G-Buffer}           & GPU          & 1.19               & 3.54                & 6.43                      \\ \hline
	                                     & CPU          & 0.05               & 0.05                & 0.06                      \\ \cline{2-5}
	\multirow{-2}{*}{\textbf{Ray Mask}}  & GPU          & 0.33               & 0.38                & 0.33                      \\ \hline
	                                     & \textbf{CPU} & \textbf{1.13}      & \textbf{2.27}       & \textbf{2.99}             \\ \cline{2-5}
	\multirow{-2}{*}{\textbf{Ray Trace}} & GPU          & 0.59               & 0.65                & 0.62                      \\ \hline
	                                     & CPU          & 0.05               & 0.06                & 0.08                      \\ \cline{2-5}
	\multirow{-2}{*}{Tile-Dilate}        & GPU          & 1.48               & 1.44                & 1.43                      \\ \hline
	                                     & CPU          & 0.03               & 0.04                & 0.04                      \\ \cline{2-5}
	\multirow{-2}{*}{PP-Composite}       & GPU          & 0.85               & 0.82                & 0.82                      \\ \hline
	                                     & CPU          & 3.14               & 4.67                & 3.48                      \\ \cline{2-5}
	\multirow{-2}{*}{Others}             & GPU          & 0.39               & 0.42                & 0.29                      \\ \hline
	                                     & CPU          & 4.66               & 7.63                & 8.39                      \\ \cline{2-5}
	\multirow{-2}{*}{Total Duration}     & GPU          & 4.83               & 7.25                & 9.92                      \\ \hline
	\multicolumn{2}{|l|}{Frame Rate}                    & 205                & 148                 & 112                       \\ \hline
\end{tabular}
\end{center}
\end{table}

Hybrid MBlur, which contains an additional ray reveal process, is expectedly slower than the adapted post-process \cite{McGuire:2012:RFP} approach that has a frame rate of 295 fps for PR. As seen in \autoref{tab:mb-profiling}, the creation of the ray mask and the ray trace pass are the major components of the ray reveal algorithm. The ray trace pass increases in duration on the CPU with the geometric complexity of each shot. As such, it is expected that the G-Buffer pass aligns with this trend as it is where deferred shading is carried out. However, the durations of the other passes appear to be independent of this geometric complexity.

\subsection{Limitations and Future Work}


Since our work is focused on the main idea of real-time hybrid MBlur, some common spatial sampling constraints in MBlur are used for simplicity considerations. These constraints include linear inter-frame motion and stable lighting within the exposure time, as well as moderate screen space velocity for a reasonable tile size. However, it is possible to accommodate nonlinear MBlur at low cost with a curve-sampling scatter approach, as demonstrated in \cite{Guertin:2015:HPN}.

Additionally, the ideal indicator for termination in our ray reveal algorithm should be the difference in velocity and not luminance. However, it is inefficient to check the velocity of a hit point during the ray tracing process. Hence, as a future improvement to our approach, we hope to incorporate GeometryIndex from DirectX Raytracing Tier 1.1, which will enable the ray tracing shader to distinguish geometries. This will be more suitable than luminance in scenes with multiple overlapping moving objects of the same luminance value.

Moreover, it is possible to improve the efficiency of our pipeline by restricting the computation of velocities to a user-defined depth for the scene, if geometry movement is localized. We also intend to integrate hybrid MBlur with other real-time hybrid rendering effects, such as depth of field \citep{Tan:2020:HDF} where we mitigate similar problems of partial occlusion from post-processing approaches.
\section{\uppercase{Conclusion}}

We present a real-time hybrid motion blur rendering technique that produces more accurate partial occlusion semi-transparencies on the silhouettes of moving foreground geometry as compared to state-of-the-art post-processing techniques. By leveraging hardware-accelerated ray tracing within a ray mask, we have achieved relatively interactive frame rates for real-time rendering in games without extensive optimization. We aim to integrate and optimize multiple hybrid rendering techniques including depth of field into our hybrid rendering engine, which will be open-sourced for the benefit of the research community and industry.

\section*{\uppercase{Acknowledgements}}
We thank \citet{Wyman:2018:IDR} for the Falcor scene file of \textsc{The Modern Living Room} (\href{https://creativecommons.org/licenses/by/3.0/}{CC BY}) as well as the NVIDIA ORCA for that of \textsc{UE4 Sun Temple} (\href{https://creativecommons.org/licenses/by-nc-sa/4.0/}{CC BY-NC-SA}) and \textsc{Amazon Lumberyard Bistro} (\href{https://creativecommons.org/licenses/by/4.0/}{CC BY}). This work is supported by the Singapore Ministry of Education Academic Research grant T1 251RES1812, “Dynamic Hybrid Real-time Rendering with Hardware Accelerated Ray-tracing and Rasterization for Interactive Applications”. 
	
\bibliographystyle{apalike}
{\small
\bibliography{hmb}}

\section*{\uppercase{Appendix}}

\subsection*{Post-process Sampling Range}

\citet{McGuire:2012:RFP} uses the dominant velocity within a neighbourhood of pixels as a heuristic. It first produces a per-exposure velocity from an inter-frame motion vector, then applies a tile-dilate pass to select the dominant velocity in the neighbourhood with the largest magnitude. Here, the per-pixel velocities are clamped at the length of each tile for the tile-dilate pass to capture the maximum velocity from all possible contributions, preventing tiling artifacts.

For the tile-dilate pass, tile-dominant velocities are first selected for tiles of length $m$ in pixel units, where $m$ is a common factor of the width and height of the image buffer resolution. Then, neighbour-dominant velocities are selected for kernels of length $n$ in tile units. For example, if we take $m$ = 40, the size of each tile is hence $40 \times 40 = 1600$ pixels. If we then dilate each tile with $n$ = 3 based on its 8 direct neighbours, our dilated tile will take the maximum pixel velocity from 9 tiles $\times$ 1600 pixels/tile = 14400 pixels in total.

Compared to approaches that use per-pixel velocities directly for gathering like \citet{Ritchie:2010:SSM}, this dominant neighbour velocity approach prevents issues such as sharp silhouettes as well as moving objects shrunk in size, as illustrated in \citet{McGuire:2012:RFP}. Following its idea, the direction of sampling at each pixel is along the dominant velocity of its neighbourhood, and the range of samples is within one magnitude of this velocity. 

\subsection*{Post-process Sample Contribution}

\citet{McGuire:2012:RFP} performs a weighted average $w$ of all corresponding samples for each target pixel.

\begin{equation}
w = w_{f} + w_{b} + w_{s}{}
\label{equ:mb-allterms}
\end{equation}
\begin{equation}
w_{f} = \saturate{$1 + z$} \times \saturate{$1 - \frac{\Delta s}{v_{s}}$}
\label{equ:mb-term1}
\end{equation}
\begin{equation}
w_{b} = \saturate{$1 - z$} \times \saturate{$1 - \frac{\Delta s}{v_{t}}$}
\label{equ:mb-term2}
\end{equation}
\begin{equation}
w_{s} = \cylinder{$v_{s}$, $\Delta s$} \times \cylinder{$v_{t}$, $\Delta s$} \times 2
\end{equation}

\noindent{where}
\begin{equation}
z = \frac{z_{t} - z_{s}}{\rm SOFT\_Z\_EXTENT}
\end{equation}
\begin{equation}
\cylinder{$v$, $s$} = 1 - \smoothstep{$0.95v$, $1.05v$, $s$}
\end{equation}

In the above formulae, $z$ represents the extent that the target pixel is deeper into the scene than its sample (negative when the pixel is shallower), while $\Delta s$ refers to the distance between the screen space positions of the target pixel and its sample. $v_{t}$ and $v_{s}$ are the screen space speeds of the target pixel and its sample respectively, while $z_{t}$ and $z_{s}$ are their camera space depths. SOFT\_Z\_EXTENT is a positive-valued scene-dependent variable as introduced in \citet{McGuire:2012:RFP} used for the classification of samples into the foreground or background of the target pixel.

\subsubsection*{Case 1: Blurry Sample in Front of Any Target}

$w_f$ considers the first case where outer blur is present: a shallower sample of speed larger or equal to its offset from the target pixel in screen space can contribute to the pixel's colour as part of the foreground. The sample contributes as foreground by covering the target pixel within the exposure time. On the other hand, since a sample deeper than the target pixel stands a higher chance to be blocked during the exposure time, the first factor in \autoref{equ:mb-term1} smoothly filters out contribution from deeper samples. Given that faster pixels also contribute less for each unit (in our case it is the length of one pixel) in their trail, a comparison between the sample's speed and its distance to the target pixel is performed as the second factor of \autoref{equ:mb-term1}. 

\subsubsection*{Case 2: Any Sample behind Blurry Target}

As for inner blur, the target pixel is moving and deeper samples within its velocity range are selected to fill the transparency left by it. Hence, for $w_{b}$, the depth comparison is done reversely to filter out shallower samples. Since \citet{Jimenez:2014:ARR} illustrates that the edge of a motion-blurred object should be a smooth gradient, a comparison between the target pixel's speed and its distance to the sample is also performed as part of the second factor of \autoref{equ:mb-term2} to assign a lower weight to further samples. 

\begin{figure}[ht]
	\centering
	\subcaptionbox{$w_{s}$ excluded}{
		\includegraphics[width=0.47\linewidth]{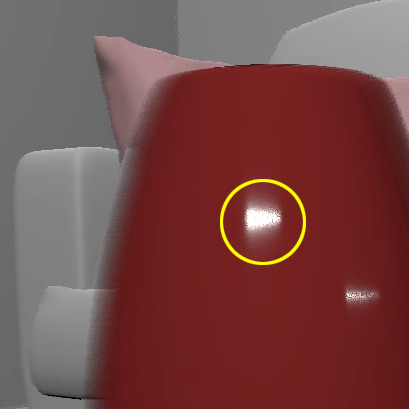}
	}
	\subcaptionbox{$w_{s}$ included}{
		\includegraphics[width=0.47\linewidth]{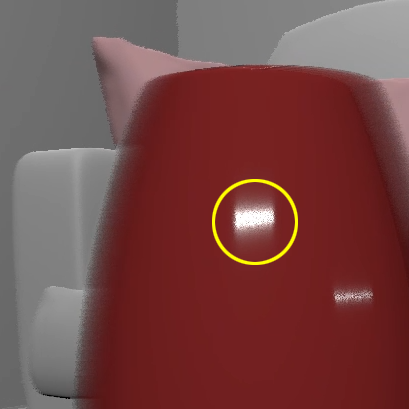}
	}
	\par\smallskip
	\caption{Specular highlight in \citet{McGuire:2012:RFP} post-processed MBlur (one-directional).}
	\label{fig:mb-term3}
\end{figure}

\subsubsection*{Case 3: Simultaneously Blurry Target and Sample}

A sample at any depth can contribute to the target pixel's colour as part of the foreground or background when both the sample and the pixel are located in each other's velocity range. $w_{s}$ blurs the target pixel and its sample into each other as a supplement to the cases not handled well by the first two terms in \autoref{equ:mb-allterms}. Since further samples have been assigned lower weights in $w_{f}$ and $w_{b}$, the area of the resulting MBlur for regions of high variance appears to shrink as the target pixel gets further from these regions. For instance, in the case of specular highlights, as shown in \autoref{fig:mb-term3}, the resulting blur appears to be triangular-shaped with the base at the exact specular highlight in the sharp rasterized image, while the desirable case should be rectangular-shaped. Since $w_{s}$ will produce a positive value when the distance between the target pixel and its sample is not significantly higher than the speed of either of them, it creates a more reasonable blur effect by enlarging the shrinking region. 

\end{document}